# First passage times of charge transport and entropy change

V. V. Ryazanov

Institute for Nuclear Research, Kiev, Ukraine, e-mail: vryazan19@gmail.com

Highlights

- All real physical processes, including of the first-passage time, occur with changes in entropy.
- Changes in entropy should be taken into account when properly studying first-passage times.
- Any other changes in entropy for other possible processes can be taken into account.
- The procedure for recording the average first-passage time does not reflect the actual situation.
- The influence of entropy changes on the average *FPT* are shown.

All real physical processes, including of the first-passage time, occur with a change in entropy. This circumstance is not taken into account when studying the first-passage time, but is illustrated in this article using the example of electron transfer through a metallic double dot. The statistics of the first-passage time of a random process *N(t)* for electrons transferred through a metallic double dot is considered. The expressions for the average first-passage time are compared with and without taking into account the change in entropy during this time. External influences on the average value of the first-passage time are considered for the case of *DC* bias voltage.

## 1. Introduction

All real physical processes that are modeled by random processes occur with a change in entropy. This also applies to first passage time (*FPT*) processes. This circumstance is not taken into account in *FPT* studies.

*FPT* s are widely used in various fields [1, 2]. The first-pass problem occupies a prominent place in natural science, since the *FPT* is a key characteristic of the kinetics of any process. *FPT* plays an important role in many areas of physics and applied mathematics, chemistry, protein folding, and even credit risk modeling. During *FPT*, entropy changes occur in the system. They affect *FPT*. In general, the connection between *FPT* and entropy change is considered in [3]. In this work, this connection is shown using a specific example of the process of electron transfer during mesoscopic charge transfer, based on the result of the experiment. It is shown that taking this dependence into account significantly affects the average FPT value.

In [4, 5], studied random times until the moment when the transfer of electric charge through a conductor reaches a given value. The *FPT* distribution of the number of electrons transferred between an aluminum conductor and a superconductor is considered. In papers [4, 5] measure the distribution of the first-passage times for the net number of electrons transferred between two metallic islands in the Coulomb blockade regime. The experimental results are consistent with the numerical calculations carried out in accordance with theory describing the distribution of the first passage time of nonequilibrium stationary Markov processes. The results of papers [4, 5] are used in this article. This article shows that the usual procedure for recording the average *FPT* does not reflect the real situation of the presence of an entropy change behind the *FPT*. An analogy can be drawn with ideal and real gases.

The core theoretical basis and approach has been explained in the author's recent paper [3]. In [3], a statistical distribution containing independent thermodynamic variables of energy and *FPT* was introduced. The factor containing the energy corresponds to the equilibrium Gibbs distribution. The nonequilibrium part of the distribution is described by a multiplier with *FPT*. Assuming that the *FPT* distribution parameters depend on the average energy values and do not depend on the random energy variable, the partition function is factorized. The factor describing the nonequilibrium part of the partition function coincides with the Laplace transform *FPT*. The Laplace transform parameter coincides with the thermodynamic parameter conjugate to *FPT* in the statistical distribution containing independent thermodynamic variables of energy and *FPT*.

The usual procedure for determining the average value of *FPT* is to differentiate the Laplace transform of *FPT* with respect to the transform parameter, after which this parameter is set equal to zero. In the approach proposed in [3], the Laplace transform parameter is equal to the thermodynamic parameter $\gamma$, which in a nonequilibrium distribution is conjugated with the random thermodynamic parameter *FPT*, just as the inverse temperature is conjugated with the random thermodynamic parameter energy. It is assumed that this is a physical quantity proportional to entropy production and fluxes [6]. This value is associated with the change in entropy in the system over *FPT*. Therefore, the average *FPT* value is written as the result of differentiating the Laplace transform *FPT* (the nonequilibrium part of the partition function) with respect to the transformation parameter. The non-zero conjugate *FPT* thermodynamic parameter, expressed through the change in entropy per *FPT*, is substituted into the resulting expression.

This paper compares the obtained expressions for the average *FPT* with a zero value of the parameter $\gamma$, conjugate to the *FPT*, and with a non-zero value of the parameter $\gamma$, expressed in terms of the change in entropy. The expressions obtained in articles [4, 5] for the distribution $P_t(N)$ for the process to take the value $N$ at a fixed time $t$ and the first-passage time probability distribution $P_N(t)$ for a stochastic process to first reach a given value $N$ at time $t$ are used, as well as the results of the experiment. An example of a *DC* bias voltage with some experimental results of [4] is considered.

In [4], a simple analytical approximation was obtained for the *FPT* distribution. Advances in nanotechnology make it possible to carry out very precise experiments with the calculation of the transferred electrons [7, 8]. Therefore, it is possible theoretically and experimentally to study the distributions of the first-passage times and waiting times [9, 10].

The second section discusses the distributions for the process to take the value $N$ at a fixed time $t$ and the first-passage-time probability distribution. In the third section, the approach of works [3] is briefly described. In the fourth section, this approach is applied to distribution (1) describing the first-passage times for the number of electrons transferred between two metallic islands.

## 2. The distribution for the process to take the value N at a fixed time t and the first-passage-time probability distribution

The fluctuations of a stochastic process $N(t)$ are usually described in terms of the distribution $P_t(N)$ for the process to take the value $N$ at a fixed time $t$. An alternative approach is to study the first-passage-time probability distribution $P_N(t)$ for a stochastic process to first reach a given value $N$ at time $t$. In the case of $P_t(N)$ the random variable is $N$ at a fixed time $t$, and in the case of $P_N(t)$ the random variable is $t$ at a fixed time $N$. Similarly, in the thermodynamics of trajectories, $s$-ensembles and $x$-ensembles are considered [24].

In [4, 5], the probability distribution $P_N(t)$ of the first passage is studied for the process of achieving a given value of the number of electrons $N$ that have tunneling between two metal islands at time $t$. For this, the complete chronology of the events of electron transfer between two metal islands was obtained in [4, 5] (see details in [4, 5]). In [4, 5], fluctuations in the time elapsed before the transfer of electric charge through the conductor reaches the specified threshold value. For this purpose, the distribution of the first passage times was measured for the net number of electrons transferred between two metal islands in a Coulomb blockade mode. The experimental results are in excellent agreement with numerical calculations based on recent studies. A theory is used that describes the exact distributions of first passage times for any nonequilibrium stationary Markov process.

The distribution of *FPT* was obtained in [4, 5] widely used in various tasks (e. g. in queue theory [11], where such distributions describe the busy interval in queuing systems). In [4] it is noted that this is a simple analytical approximation having the same structure as the random walk result [12]. This a simple analytical expression for the distribution of the first passage time takes into account the non-Gaussian statistics of electron transport through the third cumulant of the distribution $P_t(N)$. The cumulants $P_t(N)$ are defined as $C_n = (-i)^n \partial^n \mathbb{F}(\chi)/\partial \chi^n|_{\chi=0}$, where $\mathbb{F}(\chi) = \lim_{t\to\infty} t^{-1} \ln[\sum_N e^{iN\chi} P_t(N)]$ is the cumulant generating function (*CGF*). If the observation time $t$ exceeds the time $\tau_r$ required for the system to relax to a steady state after a disturbance, the *CGF* becomes independent of time $t$. The exact solution was obtained in [13].

The density of a simple analytical expression for the distribution of the first passage time is written in [4] as

$$P_N(t) = \frac{1}{t}\left|N^*\right| e^{-\frac{C_1 C_2}{C_3}t} \left(\frac{C_2 + \sqrt{C_1 C_3}}{C_2 - \sqrt{C_1 C_3}}\right)^{N*/2} I_{\left|N^*\right|}\left(\frac{C_1\sqrt{C_2^{\,2} - C_1 C_3}}{C_3} t\right), \tag{1}$$

where $N^* = [N\sqrt{C_1/C_3}]$, $C_i$, $i$=1,2,3 are the cumulants of the distribution $P_t(N)$ of the process for the number $N$ of transferred electrons at a fixed time instant $t$, $I_n(x)$ is the modified Bessel function of the first kind, and $N$ is the threshold value for the number of transferred particles. The charge transferred by them, $e^* N^*$, $e^* = e\sqrt{C_3/C_1}$ gets as close as possible to the net charge of real electrons $eN$, $e$ is the electron charge. Here, the square brackets […] indicate the rounding function.

The average electric current $\langle I \rangle$ and current noise are expressed in terms of the cumulants $C_1$ and $C_2$, $\langle I \rangle = eC_1$, $S_I = 2\int dt \left\langle I(t)I(0) - \langle I \rangle^2 \right\rangle = 2e^2 C_2$. The $C_3$ and $C_4$ values represent the third and fourth cumulants of the distribution $P_t(N)$. In [4], it was also assumed that $C_1$, $C_3>0$, and $C^2_2>C_1C_3$. Distribution (1) was obtained in [4] from a general approach with some approximations (for sufficiently large times, an approximate form of the cumulant generating function). In [4], was noted that expression (1) is fulfilled under the conditions

$$\frac{C_1\left|C_1 C_4 - C_2 C_3\right|}{12 C_2^{\,3}}\left(\frac{N}{C_1 t} - 1\right)^2 \le 1, \qquad t >> \tau_r, \qquad |N| >> 1 \cdot \tag{2}$$

In accordance with the first condition (2), relation (1) is fulfilled better near the maximum of the distribution $P_N(t)$, at $t=N/C_1$, than in the tails of the distribution. The time required for the system to return to a stationary state after an external disturbance is defined as the relaxation time $\tau_r$.

At short times expression (1) behaves like $t^{\left|N^*\right|-1}$. When the inequality holds $\left|C_3 - C_1\right| \le C_1$, it reproduces the scaling of the exact distribution $P_N(t) \sim t^{|N|-1}$ for small values of $N$. In this case, the last of conditions (2) can be weakened. At long times Eq. (1) correctly reproduces the exponential decay $P_N(t)$ predicted by the exact theory [13], but may give an inaccurate decay rate if the first of conditions (2) are violated. Further, in the Gaussian limit $C_3 \to 0$ distribution (1) reduces to the form known from the theory of Brownian motion.

The approximate distribution (1) satisfies the fluctuation relation $\frac{P_N(t)}{P_{-N}(t)} = \left(\frac{C_2 + \sqrt{C_1 C_3}}{C_2 - \sqrt{C_1 C_3}}\right)^{[N\sqrt{C_1/C_3}]}$ which does not require the system to be in equilibrium or to exhibit detailed balance, and relies only on the conditions (2). In the limit $t \gg \tau_r$, and provided the system has a well-defined temperature $T$, one can also prove an exact fluctuation relation, $P_N(t)/P_{-N}(t) = \exp[NeV_b/k_B T]$ which is a consequence of the fluctuation theorem for the electron transport and remains valid for quantum conductors described by the Schrödinger equation.

## 3. Relationship between *FPT* and entropy change.

In [5] experimentally studied fluctuations of stochastic entropy production of the electric current in non-equilibrium steady-state conditions in an electronic double dot. In this paper, we consider the thermodynamic aspects of *FPT* moments, in particular, the relationship between the first moment and the entropy change accompanying the first achievement process.

The *FPT* moments classified in the theory of stochastic processes as stopping times and Markov moments [14, 15]. The set of events observed during the random time *FPT* $T_{\gamma x}$ (3) [15] corresponds to a set describing Markov moments. This takes into account the dependence on the history of the system. The change in the entropy of the system depends on the events occurring at this time. *FPT* (3) is a multiplicative functional of a random process $X(t)$ [14]. Dependence on the system's past is important in distribution (4), (5).

In this work, arbitrary changes in entropy taken into account, which affect the moments of the process of *FPT*. Using distribution (1) makes it possible to accurately calculate the impact on the system. Changes in entropy are expressed through entropy flows into the system from the outside and entropy production in the system. The production of entropy is equal to the product of thermodynamic forces and the thermodynamic flows created by these forces. The actions on the system are also described by thermodynamic forces and entropy changes.

The distribution of the first-passage time for the number of electrons reaching a certain threshold was obtained in [4]. In [3], the argument of the Laplace transform of the *FPT* distribution is related to the change in the entropy of the system using thermodynamic relations. The average *FPT* value, in which the value of the argument of the Laplace transform of the *FPT* distribution is assumed to be zero, does not correspond to real events in which the changes in entropy and the value of this argument are not equal to zero. The task is to determine the value of the argument of the Laplace transform corresponding to the change in entropy in the real *FPT* process of reaching a given level.

The change in entropy is expressed in terms of thermodynamic flows and conjugate thermodynamic forces. The inverse relationship is also true: *FPT* affected by entropy changes caused by the introduction or change of thermodynamic forces. It is possible to formulate and study the *FPT* control problem. System entropy and *FPT* change with changes in thermodynamic forces. The processes in the system slow down or speed up.

An external *DC* voltage $V_b$ is considered as external thermodynamic forces. Non-equilibrium fluctuations of the charge-state in a hybrid double point normal metal-superconductor in the regime of strong Coulomb blockade, subject to the influence of a time-independent bias voltage, were measured in [4, 5]. An external *DC* bias voltage, $V_b$, which causes the system to go to a non-equilibrium steady state and controls the net current through the double-dot, is applied between the two leads.

In [3, 16-19] the *FPT* considered. *FPT* is defined as the time during which the random process $X(t)$ first reaches a certain threshold $a$ (3). *FPT* is by definition equal to [14, 15]

$$T_{\gamma x} = \inf\{t : X(t) = a\},\ X(0) = x > 0\ . \qquad (3)$$

The subscript $\gamma$, emphasizing the dependence on the conjugate thermodynamic parameter, is used not to confuse the variable with the temperature $T$. There are other definitions of *FPT* [14, 15]. A distribution that contains *FPT* (“lifetime” in [19]) as an additional thermodynamic parameter is introduced in [3, 16-19]. The microscopic probability density in the extended phase space with cells ($u$, $T_\gamma$) (by extended phase space we mean the phase space, the hypersurface in the phase space containing fixed values of the energy density $u$ and $T_\gamma$ compared to the hypersurface in the phase space containing only fixed values of $u$, as in the equilibrium case), where $u$ and $T_\gamma$ are thermodynamic variables, has the form

$$\rho(z;u,T_\gamma)=\exp\{-\beta u-\gamma T_\gamma\}/Z(\beta,\gamma)\,, \tag{4}$$

where $\beta=1/T$ is the inverse temperature of the reservoir ($k_B$=1, $k_B$ is Boltzmann constant), ($z=q_1,\ldots,q_N,\ p_1,\ldots,p_N$) are dynamic variables, $q$ and $p$ are the coordinates and momenta of $N$ particles of the system, the partition function is equal

$$Z(\beta,\gamma)=\int e^{-\beta u-\gamma T_\gamma}dz=\iint du\ d\,T_\gamma\ \omega(u,T_\gamma)e^{-\beta u-\gamma T_\gamma}\,. \tag{5}$$

The equations for the *FPT* distribution are conjugate to the equations for the energy distribution, for example, the Fokker-Planck equations. Since the distributions for energy depend on $z$, the same dependence is present for the *FPT* distributions. The factor $\omega(u)$ in the case of a distribution for $u$ is replaced by $\omega(u,T_\gamma)$ [3]. If $\mu(u,T_\gamma)$ is the number of states in the phase space with parameter values less than $u$ and $T_\gamma$, then $\omega(u,T_\gamma)=d^2\mu(u,T_\gamma)/dudT_\gamma$. Moreover, $\int\omega(u,T_\gamma)dT_\gamma=\omega(u)$. The number of phase points with parameters in the interval between $u$, $u+du$; $T_\gamma$, $T_\gamma+dT_\gamma$, is $\omega(u,T_\gamma)dudT_\gamma$. The parameters $\beta$ and $\gamma$ are the Lagrange multipliers. They satisfying the following expressions for the averages:

$$\langle u\rangle=-\partial\ln Z/\partial\beta_{|\gamma}\,;\qquad\qquad \langle T_\gamma\rangle=-\partial\ln Z/\partial\gamma_{|\beta}\,. \tag{6}$$

The values of energy density $u$ and *FPT* $T_\gamma$ in expressions (4)-(6) are chosen as thermodynamic parameters. The production and flows of entropy characterize non-equilibrium processes in an open statistical system. Associated with them is the parameter $\gamma$ conjugate of *FPT*. The non-equilibrium distribution (4) converges to equilibrium Gibbs distribution at $\gamma$=0 and $\beta=\beta_0=T^{-1}_{eq}$, where $T_{eq}$ is the equilibrium temperature.

The factor $\omega(u,T_\gamma)$ is the joint probability for $u$ and $T_\gamma$, considered as the stationary probability of this process (for example, [23]). We rewrite the value $\omega(u,T_\gamma)$ in the form $\omega(u,T_\gamma)=\omega(u)\omega_1(u,T_\gamma)=\omega(u)\sum_{k=1}^{n}R_k f_{1k}(T_\gamma,u)$. It is assumed that there are $n$ classes of states in the system; $R_k$ is the probability that the system is in the $k$-th class of states, $f_{1k}(T_\gamma,u)$ is the density of the distribution of *FPT* $T_\gamma$ in this class of (ergodic) states (in the general case, $f_{1k}(T_\gamma,u)$ depends on $u$). As a physical example of such a situation (characteristic of metals, glasses, etc.), one can mention the potential of many complex physical systems. Below we restrict ourselves to the case $n=1$.

The local specific entropy $s$ corresponding to distribution (4) ($u$ is the specific internal energy) is introduced by the relation [3]

$$s=-\langle\ln\rho(z;u,T_\gamma\rangle=\beta\langle u\rangle+\gamma\langle T_\gamma\rangle+\ln Z(\beta,\gamma);\qquad ds=\beta d\langle u\rangle+\gamma d\langle T_\gamma\rangle \tag{7}$$

($s\to s/k_B$, entropy is divided by $k_B$, Boltzmann's constant). This is the entropy produced in the system. We assume that in (5) $\omega(u,T_\gamma)=\omega(u)\omega_1(u,T_\gamma),\ \ \omega_1(u,T_\gamma)\sim f(T_\gamma,u)$ [3]. Here $f(T_\gamma,u)$ is the *FPT* distribution density. Suppose that this function does not depend on the random energy $u$ (possible dependence on the average energy value). Then the variables of integration are separated. The partition function statistical sum (5) is written as the product of equilibrium and non-equilibrium factors, $Z(\beta,\gamma)=Z_\beta Z_\gamma$. The non-equilibrium part of the partition function $Z_\gamma$ is the Laplace transform of the *FPT* distribution density. For internal energy and partition function the following relations are fulfilled:

$$Z(\beta,\gamma)=Z_\beta Z_\gamma,\qquad Z_\beta=\int e^{-\beta u}\omega(u)du\,,\qquad Z_\gamma=\int_0^\infty e^{-\gamma T_\gamma}f(T_\gamma)dT_\gamma\,. \tag{8}$$

From expressions (7)-(8), we obtain an equation for determining the non-equilibrium parameter $\gamma$ conjugated to the *FPT*:

$$s = s_\gamma + s_\beta = s_\beta - \Delta = \gamma \overline{T}_\gamma + \beta \overline{u} + \ln Z = \beta \overline{u} + \ln Z_\beta - \Delta, \qquad -\Delta = s_\gamma = s - s_\beta, \quad -\Delta = s_\gamma = \beta u_\gamma + \gamma \overline{T}_\gamma + \ln Z_\gamma, \quad (9)$$

$$\overline{u} = -\frac{\partial \ln Z}{\partial \beta} = u_\beta + u_\gamma, \quad u_\beta = -\frac{\partial \ln Z_\beta}{\partial \beta}\Big|_\gamma \qquad u_\gamma = -\frac{\partial \ln Z_\gamma}{\partial \beta}\Big|_\gamma = \int_0^\infty e^{-\gamma T_\gamma} \left(\frac{\partial f(T_\gamma)}{\partial \beta}\right) dT_\gamma \frac{1}{Z_\gamma},$$

where $s_\beta = \beta \overline{u}_\beta + \ln Z_\beta$, $s_\gamma = \beta u_\gamma + \gamma T_\gamma + \ln Z_\gamma$, $Z_\beta$ is “stationary” partition function; $u_\beta$ is “stationary” energy; $s_\beta$ is “stationary” entropy. Expression (7) describes the Gibbs-Shannon entropy and is related to the production of entropy in the system. Relations (6)-(9) are applied below to distribution (1).

All real physical processes are nonequilibrium and occur with a change in entropy. These changes were not previously taken into account in *FPT* calculations. The process is accompanied by a change in the entropy of the complete system. These changes in entropy, in turn, affect the course of the process, changing its moments, since the characteristics and properties of the environment in which the process takes place will change. Entropy is closely related to the passage of time, and *FPT* is also time, a certain segment of it. Complete time consists of segments of this kind.

A technique for taking into account entropy changes in FPT processes was developed in [3], [16-19]. In [20], this technique was applied to the processes of trajectory thermodynamics. In this case, internal entropy changes and entropy flows into the system from the environment were taken into account. In [21], this technique was applied to the nuclear reactor period. In [22] was applied to radiation-enhanced diffusion.

In [3], [16-19], a distribution of the form (4) is introduced, in which *FPT* is contained as a thermodynamic parameter. Such a distribution with justification based on the dynamics of the system was also obtained in [23]. Similar distributions containing *FPT* as a thermodynamic parameter have been introduced and used in trajectory thermodynamics [24]. In [3], [16-19] it is assumed that the parameter $\gamma$, conjugate to *FPT* $T_\gamma$ in distribution (4) depends on the change in entropy. This assumption is confirmed by comparison with the nonequilibrium statistical operator method and extended nonequilibrium thermodynamics [6]. The kinetic equation for *FPT* is conjugate to the kinetic equation for energy. And if energy is the main thermodynamic parameter in equilibrium, then the same basic thermodynamic parameter in nonequilibrium states is *FPT*. This is confirmed by numerous applications of *FPT* [1, 2] and a large number of articles about *FPT*. The parameter $\gamma$, conjugate to *FPT* in distribution (4), is analogous to the inverse temperature and plays the same important role as temperature. The parameter $\gamma$ equal $x$ in [24] and equal of Laplace transform argument in (8). In [6] it is shown that the parameter $\gamma$ is related to fluxes and entropy production. The usual procedure for determining averages from the Laplace transform is to differentiate with respect to the argument of the transform and then set that argument to zero. In this case, the value of $T_0$ is obtained from (12). But if we take into account the physical meaning of this argument and consider it as some physical field conjugate to *FPT*, as assumed in distribution (4), then taking into account the dependence on this field of the non-zero value of the parameter $\gamma$, after differentiation in expressions for average values, the average changes, instead of $T_0$ in (12) we obtain the value $\overline{T}_\gamma$. This is another important step towards understanding how entropy changes should be taken into account in *FPT* calculations.

In this section, the parameter $\gamma$ is associated with the total entropy change in the system during the *FPT*. Then the moments of random variables *FPT* (mean values, variances) are expressed through the parameter $\gamma$ and the total change in entropy in the system during the *FPT* time. Let us describe the algorithm for expressing moments (6), (9) through the total change in entropy. The total change in entropy $\Delta s_{tot}$ consists of the change in the entropy of the system $\Delta s_{sys}$ (6), (9) depend on the parameter $\gamma$ and the change in entropy during the exchange with the medium $\Delta s_m$. The values of $\Delta s_{sys}$ and the change in entropy during the exchange with the medium while moving along the trajectory $\Delta s_m$ also depend on the parameter $\gamma$. Let us write down the relation $\Delta s_{tot} = \Delta s_{sys}(\gamma) + \Delta s_m(\gamma)$ (18) ($\Delta s_m = S^e$) and consider it as an equation for the parameter *γ*, depending on $\Delta s_{tot}$, *γ*($\Delta s_{tot}$). Solving this equation, we obtain *γ*($\Delta s_{tot}$) and substitute it into expressions (6), (9). Thus, the parameter *γ* is included in the expression for entropy, and this same parameter is conjugated with the random thermodynamic parameter *FPT* in distribution (4); the average values and other moments of *FPT* depend on it. With this parameter, entropy changes can be mathematically integrated into existing *FPT* structures. The algorithm is indicated above: equation (9) includes the parameter *γ*. This parameter is also included in $S^e$, where $S^e$ from (18) are the quantities responsible for the exchange of entropy with the environment. These expressions for the thermodynamics of trajectories are written out in (19)-(23). Substituting expressions for $S^e$ and $\Delta s_{sys}$ into (18), we obtain equation for *γ*($\Delta s_{tot}$). Then we substitute the resulting expression for *γ*($\Delta s_{tot}$) into the expressions for moments, for example, into formula (12) instead of *s*=*γ*, obtaining $T_\gamma[\gamma(\Delta s_{tot})]$. This article uses experimental results in calculations of $\Delta s_m = S^e$.

From expression (4) it is clear that at *γ*=0 this distribution becomes the equilibrium Gibbs distribution. The $\gamma T_\gamma$ pair in distribution (4) is responsible for nonequilibrium phenomena, and an increase in the parameter *γ* corresponds to an increase in deviation from equilibrium. In [23], a distribution of the form (4) is written for arbitrary thermodynamic parameters, and not only for energy, as in (4). In [24], instead of a pair of conjugate quantities *βu*, reciprocal temperature *β* and energy density *u* from (4), a pair *sK* appears, where *K* is the dynamic activity, the number of events on the trajectory during the time *FPT τ*, *s* is the physical field conjugate to the parameter *K*. In [6, 21], questions of the physical meaning of the parameter *γ* are raised and relations for *γ* are obtained. In (9) the difference between the nonequilibrium and equilibrium (at *γ*=0) entropy of the system $\Delta S^{sys} = s_\gamma$ is written in (18), the quantities $S^e$ responsible for the exchange of entropy with the environment are added and the total change in entropy $\Delta S_{tot}$ is obtained, which appears above and for which the dependence *γ*($\Delta s_{tot}$) is sought. The same procedure was carried out in [20].

From distribution (4) the *FPT* moments are found, for example, the average value (6). These quantities depend on the parameter *γ*. Substituting the obtained dependences *γ*($\Delta s_{tot}$), we obtain expressions for the moments associated with the change in entropy. Correspondence is established at the level of average values, although such a relationship exists at the level of random variables. Thus, each of the events of dynamic activity *K* in the thermodynamics of trajectories changes the entropy of the system. Explicit expressions for such changes were obtained in [25]. Also important are the integral fluctuation relations for the production of entropy at stopping times and the statistics of infima and the stopping times of entropy production, discussed in [26-28].

Equilibrium distributions are an idealization; equilibrium states do not exist in nature. The description of equilibrium states represents a certain important stage in the development of statistical physics. Subsequently, relatively recently, progress has been made in the description of nonequilibrium phenomena. Fluctuation relations have been developed theoretically and tested experimentally [29-32], thermodynamic uncertainty relations have been obtained [33-35], and thermodynamics of trajectories have been developed [24, 36-38]. These and other achievements are associated with the theory of random processes. This science developed in parallel with statistical physics; they complemented and mutually enriched each other [39]. And the connection between entropy changes and FPT is also largely based on the theory of random processes [26-28].

The main problem in *FPT* research is determining the *FPT* distribution and its Laplace transform. The behavior of the average *FPT* value, its increase or decrease depending on the conjugated thermodynamic parameter, depends on the type of distribution. In the thermodynamics of trajectories, the dynamics of the state of the system is determined, and the stationary eigenvalue of the kinetic operator, when using the theory of large deviations, is represented as the Laplace transform of the distribution *FPT* [24, 36, 38]. In this article, as in [40], the probability density function is determined from empirical observations and is presented by an experimentally distribution. There are other approaches to describing this important indicator of the behavior of a statistical system.

If we compare the proposed approach with existing methods of *FPT* analysis, we should note its relative simplicity. The initial distribution is type (4). From it the average values and other *FPT* moments are determined. And taking into account the influence of entropy changes over the *FPT* time on the average *FPT* values may be important in various applications of the approach, such as in the period of nuclear reactors [21], thermodynamics of trajectories [20], electronic devices, or in this paper based on experimental results.

## 4. Time elapsed until the electric charge transferred through a conductor reaches a given threshold value. Comparison of theory with experimental data

In [12, 41], an exact result was obtained for a one-dimensional biased random walk:

$$P_N(t)=\frac{1}{t}\left|N\right|e^{-(\Gamma_++\Gamma_-)t}\left(\frac{\Gamma_+}{\Gamma_-}\right)^{N/2}I_N(2\sqrt{\Gamma_+\Gamma_-t})\,. \quad (10)$$

In [4] at $\Gamma_\pm=\frac{C_1}{2C_3}(C_2\pm\sqrt{C_1C_3})$, expression (1) is written. Here $\Gamma_\pm$ are the rates of jumping forward and backward. In [42], this model describes the transport of charged particles through a voltage biased tunnel junction. In [4], from (1)-(2), approximations were obtained for short times, for small values of *N*, for large times, as in the exact theory [13, 43], for a weakly non-Gaussian random process, as well as for the Gaussian limit.

In [4], the predictions of the exact theory [13, 41, 43] were compared with experimental results. A perfect agreement was found at sufficiently long times determined by expression (2). In [4], a simple and universal approximation was also written for the *FPT* distribution (1) taking into account the non-Gaussian statistics of one-electron tunneling using the third cumulant $C_3$ of the distribution of the number of transmitted electrons.

The Laplace transform of the distribution (1) has the form

$$Z_{s(=\gamma)}=\frac{(2\Gamma_+)^{N^*}}{\left(s+\Gamma_\Sigma+\sqrt{(s+\Gamma_\Sigma)^2-4\Gamma_+\Gamma_-}\right)^{N^*}},\quad \Gamma_\Sigma=\Gamma_++\Gamma_-=\frac{C_1C_2}{C_3},\quad \Gamma_+-\Gamma_-=\frac{C_1^{3/2}}{C_3^{1/2}}, \quad (11)$$

$$\Gamma_+\Gamma_-=\frac{C_1^2}{4C_3^2}(C_2^2-C_1C_3)\,.$$

The mean value of *FPT* determined from expressions (6), (8), (11) is equal to ($s$=$\gamma$)

$$\overline{T}_\gamma=\frac{N^*}{\sqrt{(s+\Gamma_\Sigma)^2-4\Gamma_+\Gamma_-}}=\frac{T_0}{\sqrt{1+\dfrac{s(s+2\Gamma_\Sigma)}{(\Gamma_+-\Gamma_-)^2}}},\quad T_0=\overline{T}_{\gamma=0}=\frac{N^*}{\Gamma_+-\Gamma_-}=\frac{N}{C_1}. \quad (12)$$

Using relations (12), the calculations are carried out and the expressions for $\overline{T}_\gamma$, the average *FPT* for non-zero values of $\gamma(\Delta)$, the parameter $\gamma$, which depends on the change in entropy $\Delta$ from (9), and the expression for $T_0$ for zero values of $\Delta$ and $\gamma(\Delta)$ are compared with each other.

By $\gamma\geq 0$, $\overline{T}_\gamma\leq T_0$. We seek the dependence $\gamma(\Delta)$ from equation (9). Expressions for $\Gamma_+$ and $\Gamma_-$ are written in [4, 5]. So, the expression for $\Gamma_+$ has the form

$$\Gamma_+=\Gamma_{L\to R}(\Delta E)=\frac{1}{e^2R}\int dE n_L(E)n_R(E-\Delta E)f_L(E)(1-f_R(E-\Delta E))\,, \quad (13)$$

where $\Delta E = -eV_b$. Here $R=R_{nm}$ is the resistance of the transition in which the electron jump occurs, $n_i(E)$ and $f_i(E)$ are the density of states and the distribution function in the initial electrode, $n_f(E)$ and $f_f(E)$ are the density of states and the distribution function in target electrode (is Fermi function). The density of states in superconductors has the usual form $n_S(E) = \theta(|E| - \Delta_1)|E| / \sqrt{E^2 - \Delta_1^2}$, where $\Delta_1$ is the superconducting gap, and in normal metals it is equal to *1*. The corresponding rates are obtained in [44-46].

The rates can be conveniently expressed in the form [47]:

$$\Gamma_+ = \Gamma_d e^{-\beta eV_b/2}, \qquad \Gamma_- = \Gamma_d e^{\beta eV_b/2}. \tag{14}$$

In accordance with the general approach to diffusion in an external field [48], we write the value $\Gamma_d$ in the form $\Gamma_d = \Gamma_0 e^{-\beta E_a}$, where $E_a$ is the maximum potential energy between the "electron islands". In [5] and [49] the expression for electrostatic energy in an electronic double dot is written. In case $V_b$=0

$$E_{dot}(n) = \frac{E_{C_1}}{2}(n_L - n_{g,L})^2 + \frac{E_{C_2}}{2}(n_R - n_{g,R})^2 + E_{C,m}(n_R - n_{g,R})(n_L - n_{g,L}), \tag{15}$$

where $E_{C_1}, E_{C_2}$ are the charging energies of the islands, $E_{Cm}$ is the electrostatic coupling energy, $n_L$, $n_R$, $n_{g,L}$, $n_{g,R}$ are the charge and gates charge states of the left and right dots.

For simplicity, let's take the values $V_{gL} = V_{gR} = 0,\ n_{gL} = n_{gR} = 0,\ n_L = n_R = 1/2$. From [5] we take the values $E_{CL} = E_{C_1} = 60\ \mu eV,\ E_{CR} = E_{C_2} = 40\ \mu eV,\ E_{C,m} = 10\ \mu eV$. Then from (15) we obtain $E_{dot} = 15\ \mu eV$, $T_{eff} = 1{,}175\ K,\ \beta^{-1} = k_B T_{eff} = 1{,}6215*10^{-23} J,\ \beta E_a = 0{,}148$. Value $T_{eff} = 1{,}175\ K$ taken from Table 1 [5] as an average $T_{eff}$ for voltages $V_b = 25\ \mu V$ and $V_b = -25\ \mu V$.

In the course of reaching the boundary *N* by the random process *N(t)*, the entropy of the system [5] changes by the value $\Delta=\Delta_e$. These changes must be taken into account. Therefore, one cannot assume in (12) $\gamma=0$, considering the value of $T_0$ as the average value of the time of the first achievement in the absence of impacts on the system. An internal change in entropy $\Delta_e$ occurs and in the absence of any external influences in the system. The value of $\Delta_e$ can be determined, for example, from the relations of extended irreversible thermodynamics [50] $\Delta_e = \tau_e ii/2\rho\sigma_e T$, where $i \sim V_b$ is electric flux (current), $\tau_e$ is relaxation time of currents, $\sigma_e$ is electrical conductivity, or from results [5].

If there are other processes in the system that cause the corresponding changes in entropy, then $\Delta = \Delta_e + \Delta_{ext}$, where $\Delta_{ext}$ are the changes in entropy associated with other physical processes, for example, heat conduction.

If we write relation (9) using relations (6)-(8) for the nonequilibrium partition function (11) and expression (12), we obtain a transcendental equation for the parameter $\gamma$ conjugate to *FPT* $T_\gamma$ in distribution (4). We want to express this parameter through the change in entropy $-\Delta = s_\gamma$, substitute it into expression (12) and obtain the connection between the average *FPT* $\overline{T}_\gamma$ and the change in entropy.

To write down an explicit solution to the transcendental equation, we expand the expressions included in it into a series to quadratic terms in $\gamma$. This expansion for expressions (11)-(12) has the form

$$\ln Z_\gamma \approx -\frac{N^*}{(\Gamma_+ - \Gamma_-)}[\gamma - \gamma^2 \frac{(\Gamma_+ + \Gamma_-)}{2(\Gamma_+ - \Gamma_-)^2}], \quad \gamma\overline{T}_\gamma \approx \frac{N^*}{(\Gamma_+ - \Gamma_-)}[\gamma - \gamma^2 \frac{(\Gamma_+ + \Gamma_-)}{2(\Gamma_+ - \Gamma_-)^2}], \quad \ln Z_\gamma + \gamma\overline{T}_\gamma \approx 0.$$

Equation (9) takes the form $-\Delta = s_\gamma = \beta u_\gamma$. What remains is the expression for $\beta u_\gamma$, the expansion for which has the form

$$\beta u_\gamma \approx \gamma^2 a + \gamma b, \tag{16}$$

$$a=-N^{*}(1.085+\beta eV_b/2-\beta E_a)\frac{(\Gamma_{+}+\Gamma_{-})}{2(\Gamma_{+}-\Gamma_{-})^{3}}-\frac{N^{*}}{2(\Gamma_{+}+\Gamma_{-})^{4}}[(\Gamma_{+}-\Gamma_{-})\beta\frac{\partial(\Gamma_{+}+\Gamma_{-})}{\partial\beta}-3(\Gamma_{+}+\Gamma_{-})\beta\frac{\partial(\Gamma_{+}-\Gamma_{-})}{\partial\beta}],$$

$$b=N^{*}(1.085+\beta eV_b/2-\beta E_a)\frac{1}{(\Gamma_{+}-\Gamma_{-})}-N^{*}\frac{1}{(\Gamma_{+}-\Gamma_{-})^{2}}\beta\frac{\partial(\Gamma_{+}-\Gamma_{-})}{\partial\beta}.$$

In (16), the relation that follows from (14) is used: $\frac{\partial\Gamma_{\pm}}{\partial\beta}=\Gamma_{\pm}(-E_a\mp eV_b/2)$, as well as the relation $\partial N^{*}/\partial\beta=N^{*}(-E_a+eV_b/2+1.085/\beta)$, where for the derivative with respect to $\beta$ of the cumulant $C_1$, proportional to the average current, the dependence of the electronic conductivity is taken in the form indicated in [51]. The expression for $\beta$ uses the temperature $T_{eff}$ from Table 1 in [5].

The solution to equation $-\Delta=s_{\gamma}=\beta u_{\gamma}$ with function (16) has the form

$$\gamma=(-b\pm\sqrt{b^{2}-4as_{\gamma}})/2a. \qquad (17)$$

We select the "-" sign in (17), at which $\gamma_{|\Delta=0}=0$. When choosing the "+" sign, we obtain a stationary nonequilibrium state, which is realized at small negative values of $\gamma$.

Substituting this parameter $\gamma$ (17) into an expression (12) makes it possible to find the value of the average *FPT* $\overline{T}_{\gamma}$ at $s=\gamma$, obtained from distribution (4)-(6), depending on the change in entropy $\Delta_e$ in accordance with Eq. (9). Solution of equation (17) satisfy the condition $\gamma_{|\Delta=0}=0$. We use the branch of solution (17) that corresponds to the positive values of the root expression in the denominator (12).

In relation (9) the value of -Δ is equal to the entropy changes within the system $\Delta S^{sys}=s_{\gamma}$. Let us determine this value using the relation

$$S_{tot}=\Delta S^{sys}+S^{e}, \qquad (18)$$

where $S_{tot}$ is the total change in entropy, $S^{e}$ are the quantities responsible for the exchange of entropy with the environment. This value $S^{e}$ was calculated using the expressions and data defined for the problem under consideration in [5]. In [5] is written "If the environment consists of several thermal reservoirs and local detailed balance holds, the mesoscopic entropy flow to these reservoirs $S^{e}(t)=-\Sigma_k Q_k(t)/T_k$, where $-Q_k(t)$ is the heat dissipated to a thermal reservoir at temperature $T_k$". In [5] expressions for $Q_k(t)$ are written for the situation under consideration, and also have used the definition of the entropy flow $S^{e}(t)=log\Pi_{j=1}^{N(t)}\Gamma(n_{j-1}\to n_j)/\Gamma(n_j\to n_{j-1})$. Transition rates between different charge states of the double dot $\Gamma(n_{j-1}\to n_j)$ are given in [5]. The empirical transition rates are calculated using Eq. (D1) in [5], $\Gamma(n\to n`)=N_{n\to n`}/P^{st}(n)\tau$, $N_{n\to n`}$ is the number of transitions from state $n$ to state $n`$, $n\to n`$. All the data is obtained from counting statistics of experimental traces of durations of at least 1 hour. To obtain the stationary transition rates $\Gamma(n\to n`)$ from the time trace $\{n_{tot}(t)\}$, in [5] (Supplement D) count the number of transitions $N_{n\to n`}$ that occur from state $n$ to state $n`$ for each bias voltage $V_b$ value. In [5] calculate the transition rate between the states $n$ and $n`$ using [52] (D1) where $\tau$ is the time duration of the experiment and $P^{st}(n)=\tau_n/\tau$, (D2) [5] is the empirical steady-state occupation probability of the state $n$, calculated as the fraction of the total time when the double dot stays in state $n$. The traces of stochastic entropy production $s(t)=S_{tot}$ and of the entropy flow $S^{e}(t)$ are calculated using the empirical transitions rates (D1), occupation probabilities $P^{st}{}_{n}$ (D2) from the time trace $\{n_{tot}(t)\}$. The expression (C10) $-\tilde{Q}_{m}^{n}/T_{el}=\ln\tilde{\Gamma}_{m}^{n}/\tilde{\Gamma}_{n}^{m}$ in [5], where the electronic temperature of the superconducting and normal-metal components $T_{el}\approx 170$ mK, as well as the values of $T_{eff}$ from Table 1 in [5], are taken into account. The values $S_{tot}$ are found from Table 2 in [5], they correspond to equation (1) in [5]. From (18) we find $\Delta S^{sys}$ (the last line in Table 1), we replace in expressions (16) - (17) - $\Delta = S^{sys}$ with $\Delta S^{sys}$ and, using these data, we build Fig. 1-2.

Following [20], we will show how quantity $S^{e}$ is defined in the thermodynamics of trajectories. The change in entropy during the exchange with the medium while moving along the trajectory $\boldsymbol{Y}_K(C_0 \to C_1 \to ... \to C_K)$ is ([20], where $\Delta s_m$ equal $S^e$, $C_j$ is the configuration in moment *j*)

$$\Delta s_m[C(t)] = \sum\nolimits_{\alpha=1}^{K} \Delta s_1(C_{\alpha-1}, C_\alpha), \tag{19}$$

as dynamic activity $K$ changes from $K_0$ to $K_\gamma$,

$$\Delta s_{m(K_0 K_\gamma)}[C(t)] = \sum\nolimits_{\alpha=K_0}^{K_\gamma} \Delta s_1(C_{\alpha-1}, C_\alpha), \tag{20}$$

where we sum over all configuration changes. The corresponding dynamical partition function is [53, 54]

$$Z(\lambda,\tau) = \left\langle e^{-\lambda s_m} \right\rangle \sim e^{\tau\theta(\lambda)}, \tag{21}$$

where $\lambda$ is the parameter conjugate to $s_m$., $\theta(\lambda)=g^{-1}(\lambda)$ [20, 24], the function $g(s=\lambda)=K^{-1}lnZ_{s=\lambda}$, $Z_{s=\lambda}$ from (11), $\tau$ is the observation time. In this case, the theory of large deviations was used, large values of $K$ and *FPT* $\tau$ were selected. In this article, the value $S^{e}$ was calculated using the expressions and data defined for the problem under consideration in [5]. In expression (11) and in subsequent calculations, the values of $N$ and *FPT* can be arbitrary (with some restrictions). In analogy with the activity, the mean entropy production rate in the $\lambda$-ensemble is given $\left\langle s_m \right\rangle / \tau = -\partial\theta(\lambda)/\partial\lambda$ . Assuming that the quantity $\Delta s_1 = \Delta s_1(C_{\alpha-1}, C_\alpha)$ is constant, expression (20) takes the form

$$\Delta s_{m(K_0 K_\gamma)}[C(t)] = \Delta s_1(\left\langle K_\gamma \right\rangle - \left\langle K_{\gamma=0} \right\rangle). \tag{22}$$

Using the results of [55] (19), we obtain for the term from (38), when

$$\left\langle s_m \right\rangle = -\tau\partial\theta(\lambda)/\partial\lambda, \quad \tau=\tau_0, \; \Delta\left\langle s_m \right\rangle = \left\langle s_{m|K_\gamma} \right\rangle - \left\langle s_{m|K_0} \right\rangle = -\tau[\partial\theta(\lambda)/\partial\lambda\big|_{\lambda=\gamma} - \partial\theta(\lambda)/\partial\lambda\big|_{\lambda=0}], \tag{23}$$

Expression (23) coincides with (22) at $\Delta s_1$=(+/-)1 [26].

Expressions (19)-(21) were obtained in [54] for the thermodynamics of trajectories, expanding the *s*-ensemble approach to driven systems based on the results obtained in [53].

The values of $\Delta_e = \Delta S^{sys} = s_\gamma$ change in entropy at different voltages are taken from Eq (18), Fig. 4, 12 and Table II from [5]. Fig. 1 shows the dependences $T_0(N)=N/C_1$ (blue dashed line) and $\overline{T}_\gamma(\Delta_e) = \overline{T}_{\gamma(\Delta_e)}$ (red full line) for different values of $N$ at bias voltage $V_b$=90 $\mu V$. Let us consider external influences using the example of the applied bias voltage $V_b$. Fig. 2 shows the dependence of $\overline{T}_\gamma(\Delta_e)$ and $T_0$ on the applied voltage $V_b$ at the level $N$=10.

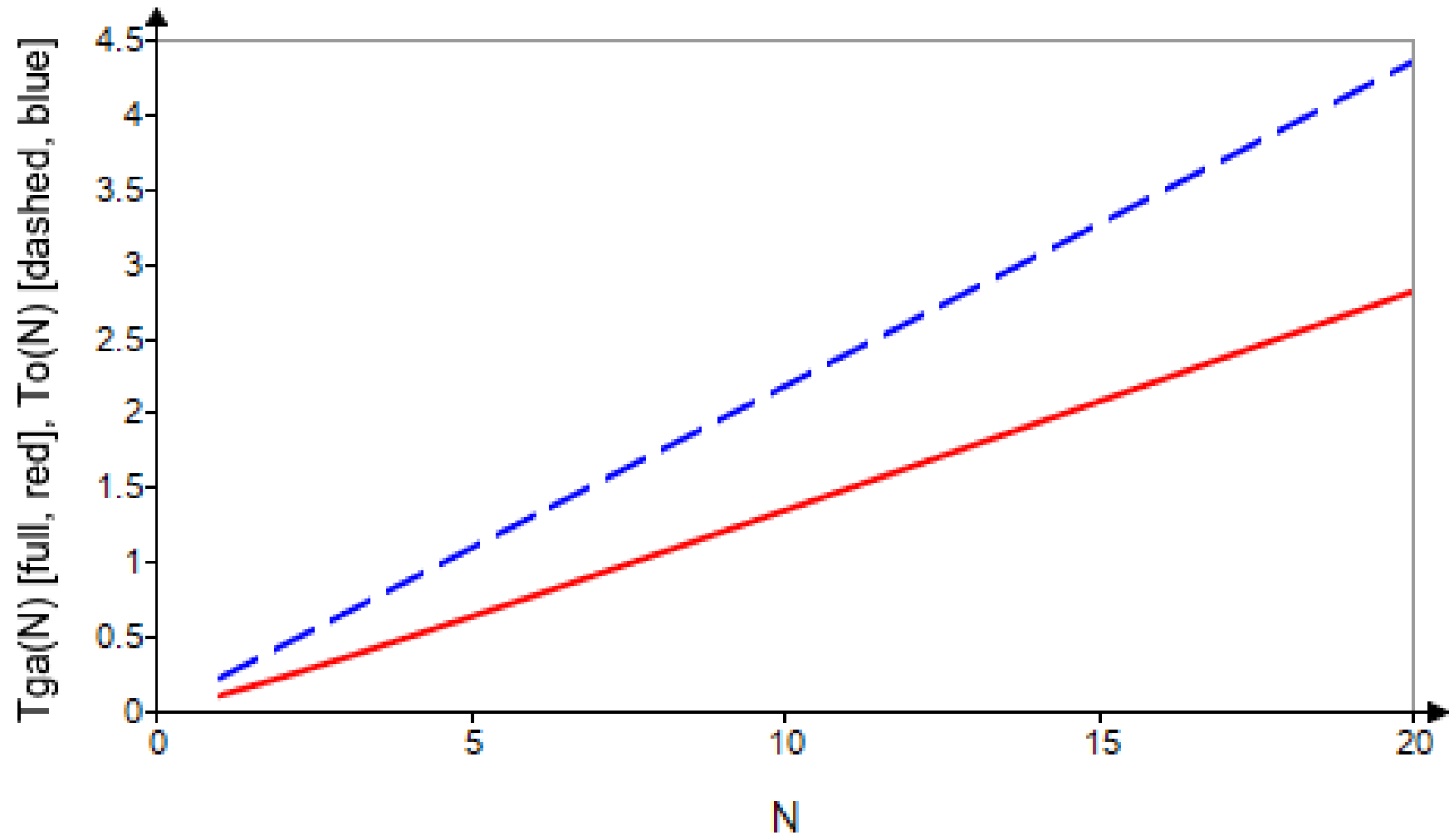

**Fig. 1**. Dependencies $T_0(N)=\overline{T}_{\gamma=0}=N/C_1$ (blue dashed line) and $Tga(N)=\overline{T}_{\gamma(\Delta_e)})=\overline{T}_{\gamma(\Delta)}$ (red full line approximating the calculated points) for the value $N$ of the process $N(t)$ to take at a fixed time $t$. Regarding the random process of achieving a given value $N$, $N(t)$ is the net number of transmitted electrons [4, 5]. Bias voltage $V_b$=90 μV.

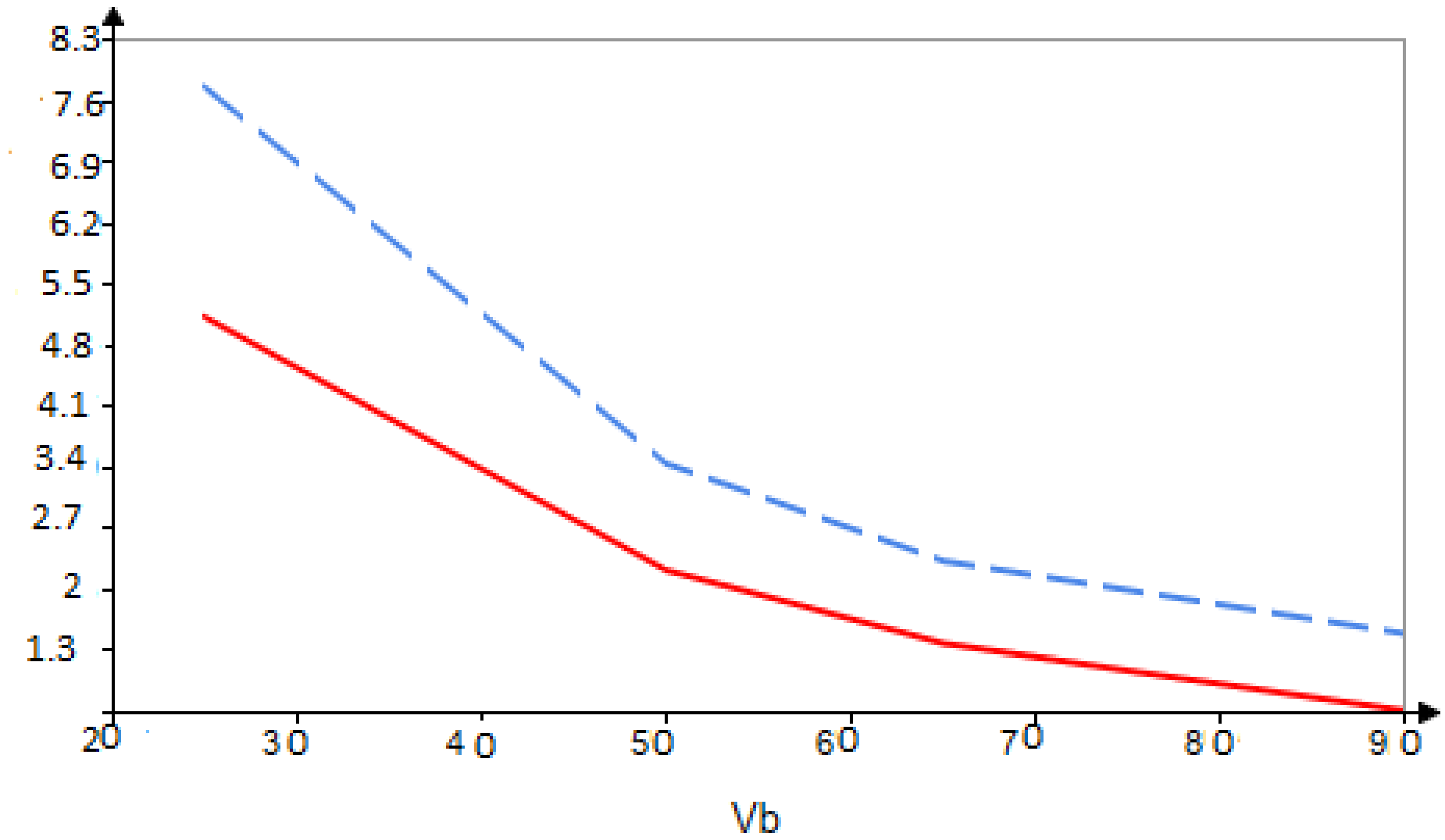

**Fig.2**. Calculated from relations (7)-(18) and Table 1 dependences of the average time $Tga(V=V_b)=\overline{T}_{\gamma(\Delta_e)}$ (red full line) and $T_0(V=V_b)=\overline{T}_{\gamma=0}=N/C_1$ (blue dashed line) to reach the level $N$=10 for different values of $V=V_b$ (25, 50, 65, 90 μV). The points in Fig. 2 calculated according to Table 1. Tga($V_b$) [full, red], $T_0$ [full, blue].

The figure is symmetrical about the $y$-axis for negative $V_b$ values.

The data used in the calculations are summarized in Table 1.

Table 1.

| $V_b$ $\mu V$ | 90 | 65 | 50 | 25 |
|---|---|---|---|---|
| $-\beta eV_b/2$ | 0.522 | 0.32 | 0.23 | 0.093 |
| $C_1$ $Hz$ | 4.6 | 3,25 | 2.48 | 1.23 |
| $C_2$ $Hz$ | 9.27 | 8.725 | 8.48 | 8.23 |
| $C_3=\alpha^2 C_1$ $Hz$ | 2.18 | 1.54 | 1.18 | 0.583 |
| $\Gamma_+$ $Hz$ | 13.12 | 11.54 | 10.7 | 9.82 |
| $\Gamma_-$ $Hz$ | 6.439 | 6.846 | 7.1 | 7.55 |
| $T_0$ $c$ | 2.17 | 3.077 | 4.037 | 8.13 |

| $\bar{T}_{\gamma(\Delta_e)}$ *c*, N=10 | 1.341 | 2.067 | 2.866 | 5.629 |
|---|---|---|---|---|
| $S_{tot}$ | 1.83 | 0.95 | 0.6 | 0.31 |
| $\Delta S^{sys}$ | 1.79 | 0.8 | 0.41 | 0.16 |

Table 1. Used in the calculations of Fig. 2, the data obtained from relations (13) - (15) and expressions for the cumulant *Ci=1,2,3* [4, 5].

## 5. Conclusion

The main purpose of this article: to show the need to take into account the change in entropy in *FPT*. The average *FPT* is associated with changes in entropy that occur during this process. The effect of such changes on the average *FPT* is shown using electrons transferred through a metallic double dot in the Coulomb-blockade as an example. The proposed approach makes it possible to take into account any other changes in entropy that correspond to other possible processes occurring in the system. The mean *FPT* values in accordance with expression (12) decrease when the system is influenced (for this process, $\gamma \geq 0$ and distribution (1)).

It is shown that the average *FTP* at zero value of the Laplace transform argument of the density distribution of the *FTP*, which is usually used to determine the average value of the *FTP*, does not reflect the influence of real processes on the average *FTP*. In Fig. 1 shows how taking into account the changes in entropy accompanying the process of the *FPT* affects the average value of *FPT*. In Fig. 2 shows how the applied voltage affects the system.

In this paper, the possibilities of taking into account the influence of entropy changes on the average *FPT* are shown. This is done on the basis of experimental results. It can be shown that this effect is different for different physical systems. For example, for neutrons in a nuclear reactor, the effect of changes in entropy on the reactor period during this period not so great for almost all values of the multiplication factor (with some exceptions) [21]. The influence of external influences is shown on the example of voltage affects. However, the effects can be very diverse (temperature effects, mechanical, etc.). The approach proposed in [3] opens up opportunities for the study of arbitrary influences by including them in changes in entropy. There are also opportunities to control the *FTP*, as noted in [3].

The example of electron transfer transferred between two metallic islands in the Coulomb blockade regime considered in the article demonstrates the significant influence of taking into account the change in entropy on the speed of electron transfer when the number of transferred electrons reaches a certain specified value. In all electronic devices, electron transfer plays a major role. Therefore, the results obtained may be important and useful in the description and calculations of materials, devices, circuits and systems.

Data Availability Statement: The datasets generated during and/or analysed during the current study are available in the arxiv.org/abs/2201.06497 repository.